\pdfoutput=1

\documentclass[conference]{IEEEtran}

\usepackage{times}
\usepackage{latexsym}
\usepackage{hyperref}
\usepackage{cite}
\usepackage{float}
\usepackage{amsmath}
\usepackage{amssymb}

\usepackage[T1]{fontenc}
\usepackage[utf8]{inputenc}

\usepackage{microtype}
\usepackage{inconsolata}
\usepackage{graphicx}
\usepackage{caption}
\usepackage{subcaption}

\hypersetup{bookmarks=false}
\hypersetup{hidelinks}

\begin{document}
\title{Cardi-GPT: An Expert ECG-Record Processing Chatbot}

\author{
\IEEEauthorblockN{Koustav Mallick, Neel Singh, Mohammedreza Hajiarbabi}
\IEEEauthorblockA{\\
    Department of Computer Science \\
    Purdue University Fort Wayne
}
}

\maketitle
\begin{abstract}
Interpreting and communicating electrocardiogram (ECG) findings are crucial yet challenging tasks in cardiovascular diagnosis, traditionally requiring significant expertise and precise clinical communication. This paper introduces Cardi-GPT, an advanced expert system designed to streamline ECG interpretation and enhance clinical communication through deep learning and natural language interaction. Cardi-GPT employs a 16-residual-block convolutional neural network (CNN) to process 12-lead ECG data, achieving a weighted accuracy of 0.6194 across 24 cardiac conditions. A novel fuzzification layer converts complex numerical outputs into clinically meaningful linguistic categories, while an integrated chatbot interface facilitates intuitive exploration of diagnostic insights and seamless communication between healthcare providers. 

The system was evaluated on a diverse dataset spanning six hospitals across four countries, demonstrating superior performance compared to baseline models. Additionally, Cardi-GPT achieved an impressive overall response quality score of 73\%, assessed using a comprehensive evaluation framework that measures coverage, grounding, and coherence. By bridging the gap between intricate ECG data interpretation and actionable clinical insights, Cardi-GPT represents a transformative innovation in cardiovascular healthcare, promising to improve diagnostic accuracy, clinical workflows, and patient outcomes across diverse medical settings.
\end{abstract}
\section{Introduction}
Electrocardiography (ECG) is an indispensable tool in diagnosing and monitoring cardiovascular diseases, with the 12-lead ECG configuration offering a comprehensive view of cardiac electrical activity. Its widespread use and significant clinical value make accurate and timely interpretation crucial for effective patient care \cite{esc_ai_ecg_diagnostics}. Manual ECG interpretation, while reliable, is restricted by expert availability, time-intensive, error-prone, and challenged by growing patient data complexity.\cite{salerno2003ecg_competency}. These challenges are further exacerbated in remote areas where access to doctors is limited, but paramedical staff are available. In such settings, an intelligent expert system, incorporating knowledge from medical professionals, can enable paramedics to assess disease severity and make informed decisions, including referring patients to the nearest medical center where doctors are available. Automated systems that integrate predictive modeling and expert knowledge have demonstrated significant potential to address these challenges. Predictive models, often leveraging machine learning algorithms, analyze vast datasets to identify patterns and enhance diagnostic accuracy \cite{ose2024ai_ecg_review}, while expert systems emulate human expertise through rule-based logic, providing consistent and explainable solutions. This dual approach holds potential for transforming diagnostic workflows, especially in resource-constrained environments.

\section{Background}

\subsection{Electrocardiogram}
An electrocardiogram (ECG/EKG) is a diagnostic tool that records the heart's electrical activity over time through electrodes placed on the body. The standard 12-lead configuration provides a comprehensive view of the heart’s electrical activity from multiple angles, making it invaluable for identifying cardiac abnormalities such as arrhythmias, myocardial infarction, and other cardiovascular conditions \cite{Sattar2023Electrocardiogram}. ECG data are typically represented as time series signals, with each lead capturing the activity of a specific part of the heart. Interpretation of  these data requires specialized knowledge to recognize subtle patterns and deviations across multiple leads.
\begin{figure}[h]
\centering
\includegraphics[width=0.6\linewidth]{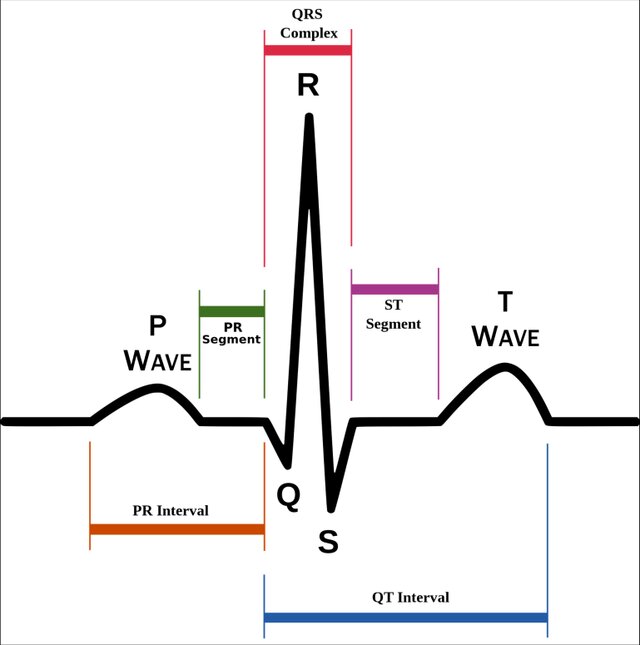}
\caption{ECG Signal.}
\label{Fig1:ECG Signal}
\end{figure}

\subsection{Expert Systems}
Expert systems are computer-based tools designed to replicate the decision-making capabilities of human experts by combining data-driven algorithms with domain-specific knowledge. Traditional expert systems in ECG analysis are often rule-based, relying on predefined if-then logic to identify specific conditions\cite{app122312342}.These systems are limited by the complexity of encoding all possible scenarios into rule sets.

\subsection{Neural Networks}
Neural networks, particularly deep learning models, have emerged as powerful tools for time series analysis, making them well-suited for interpreting ECG data. Convolutional Neural Networks (CNNs) and their variants have demonstrated high accuracy in single-lead ECG classification tasks\cite{9176396}. Recent research emphasizes the potential of neural networks as expert systems, especially when trained on expert-labeled datasets. By combining neural networks with explainable AI techniques, these systems can provide accurate predictions and insights into the decision-making process\cite{grabowski2022classificationselfsupervisedregressionarrhythmic}.

\section{Related Works}
\subsection{Rule-Based Expert Systems}
Early attempts to automate ECG interpretation relied heavily on rule-based expert systems. These systems encoded diagnostic rules provided by domain experts to classify ECG signals into specific cardiac conditions\cite{app122312342}. For instance, simple rules might identify arrhythmia based on predefined thresholds for heart rate variability or waveform amplitude. Although effective in well-defined cases, traditional systems lacked adaptability, particularly when handling edge cases or noisy datasets.
\subsection{Deep Learning in ECG Classification}
Deep learning has revolutionized ECG interpretation by enabling models to learn directly from data. More recently, deep learning approaches have become the standard, with Convolutional Neural Networks (CNNs) achieving state-of-the-art results in single-lead ECG classification tasks\cite{xiaolin2020heartbeat_classification}. The transition from single-lead to 12-lead ECG analysis introduces additional complexity due to the spatial relationships between leads. To tackle these complexities, recent studies have focused on advanced architectures like residual networks and attention mechanisms. CNNs have shown great promise due to its ability to automatic feature extraction and robustness to spatial invariance.
\subsection{Previous Works}
Several Works have shown good results. \cite{grabowski2022classificationselfsupervisedregressionarrhythmic} have achieved an accuracy of overall accuracy of 87.33\% over MIT-BIH Arrhythmia database. Zhang et. al. \cite{zhang2021interpretable} have achieved an F1 Score of $0.813$ over CPSC-2018 dataset \cite{liu2018open}. Zhao et. al. \cite{zhao2020adaptive} have achieved a 5-fold cross-validation score of 0.684 over a compiled dataset by PhysioNet\cite{alday2020classification}. On a same dataset, Bjorn\cite{singstad2020convolutional} ran another model with a combinaiton of CNN and Rule-based system to achieve 0.54. These results set the benchmark for our experiment.

\section{Methodology}
Our proposed system, Cardi-GPT, integrates predictive modeling, fuzzification, and chatbot-based solution delivery to streamline the interpretation of 12-lead ECG data.

\subsection{Dataset}
We used the PhysioNet/Computing in Cardiology dataset \cite{alday2020classification}, consisting of 43,101 12-lead ECG recordings labeled with 27 diagnoses, collected from six hospitals across four countries spanning three continents. This demographic diversity addresses a key limitation of previous works. Three diagnosis pairs, having similar clinical implications but differing regional names, were merged, reducing the total classes to 24.

Each recording is annotated with one of these 24 conditions. The dataset is imbalanced, with some conditions significantly underrepresented, and was de-noised for consistency. The recordings vary in frequency (257 Hz – 1000 Hz) and duration (6s – 60s), with a small subset (n=74) extending to ~30 minutes. Each ECG entry includes patient age and gender.

The dataset was stratified into training (72\%), validation (18\%), and test (10\%) sets to ensure proportional class representation. The training set was used for model learning, the validation set for hyperparameter tuning and performance tracking, and the test set for final evaluation to assess generalization.

\subsection{Predictive Model}
The predictive model serves as the backbone of the system, analyzing 12-lead ECG data to classify cardiac conditions.

\subsubsection{Preprocessing}
We preprocessed ECG data for standardization and reliable predictive analysis. First, All ECGs were resampled to the minimum frequency of 257 Hz. Then, signals were zero-padded to a uniform length of 4,096 samples, i.e., just under 16 seconds to accommodate the fixed input dimensions required by the convolutional neural network, preserving signal integrity. Likewise, we used a window with the length 4096 to randomly clip longer duration signals. Amplitude normalization was applied to mitigate variations caused by patient-specific factors, electrode placement, or device calibration, ensuring the model focuses on diagnostic patterns rather than irrelevant artifacts. Additionally, we scaled age into the range [0, 1] using maximum age as 100. Both age and gender were encoded using one-hot encoding, with two additional mask variables to represent missing values. These steps collectively prepared the dataset for robust and clinically relevant classification.

\subsubsection{Model Architecture}
\begin{figure}[H]
\centering
\begin{subfigure}[t]{0.45\linewidth}
    \centering
    \includegraphics[width=\linewidth]{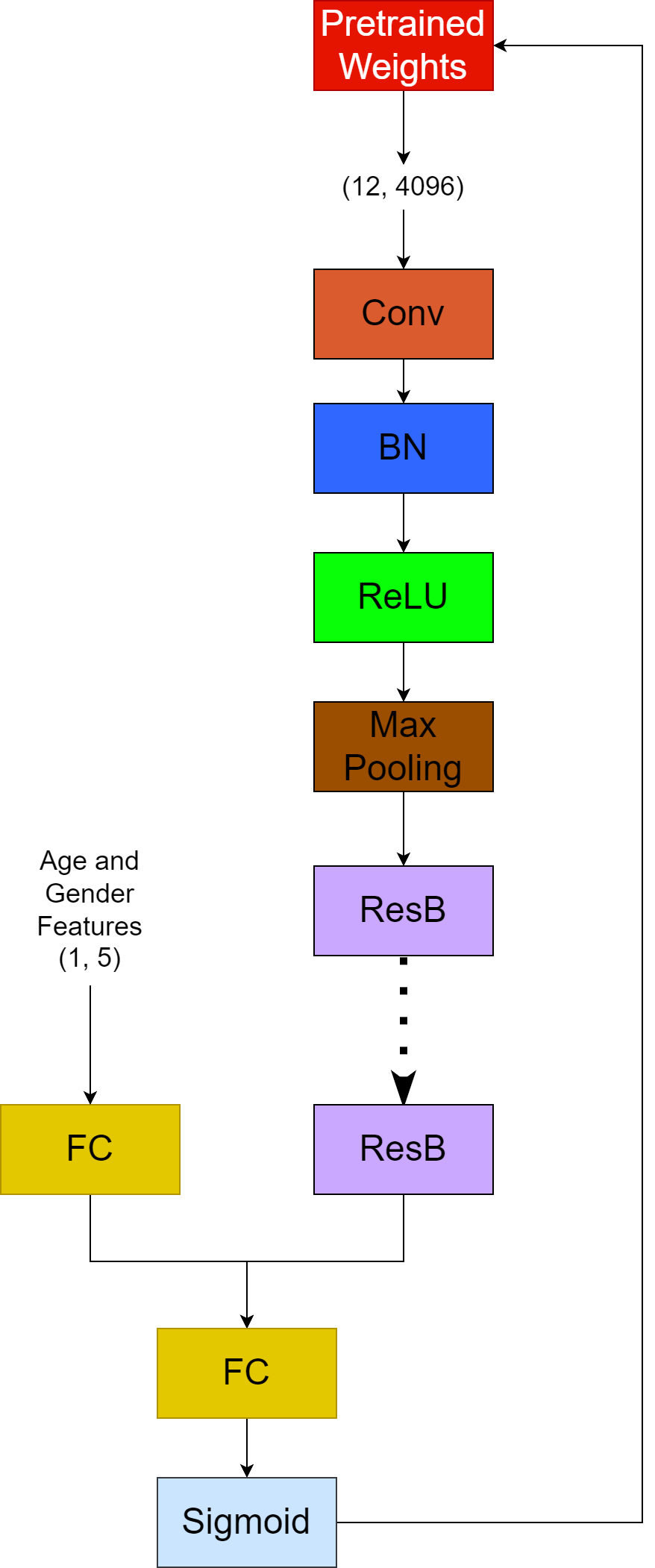}
    \caption{Predictive Model Outline}
    \label{Fig2:ModelArchitecture}
\end{subfigure}
\hfill
\begin{subfigure}[t]{0.50\linewidth}
    \centering
    \includegraphics[width=\linewidth]{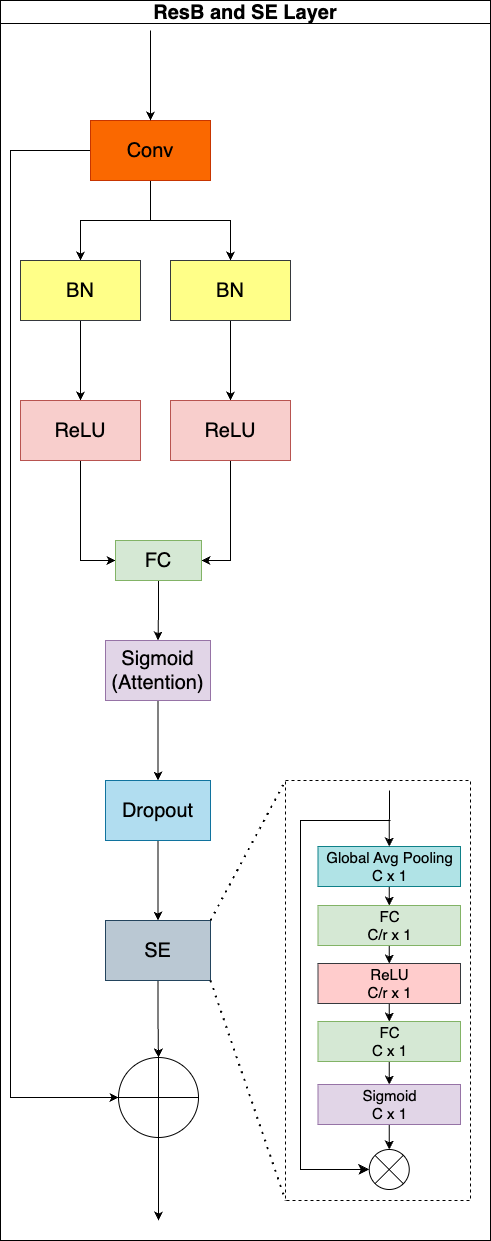}
    \caption{Expanded View of the Residual Block}
    \label{Fig3:ResidualBlock}
\end{subfigure}
\caption{Model Architecture}
\label{Fig:ModelArch}
\end{figure}

The model utilizes a 1-D Convolutional Neural Network (CNN) architecture with 16 residual blocks. Each block contains two parallel convolutional layers and  employs an attention-based sigmoid activation function to handle the multi-label nature of the classification task. Due to the fact that some samples have multiple classes of 27 clinical diagnoses, instead of using the softmax function in traditional classification problem, we assumed that each class was independent and used the sigmoid function for each output neuron to cope with this multi-task problem. The model also contains a squeeze-and-excite layer \cite{hu2019squeezeandexcitationnetworks} in order to enforce the model to facilitate a safe output (to show positive when it is very positive about the classification). A dropout rate of 0.2 is used throughout the network for regularization. Fig-\ref{Fig:ModelArch} shows a broad diagram of the architecture employed.

The first (convolutional) layer and the initial two ResBs
units have 64 convolution filters. The number of filters increases by a factor of two for every second ResB unit. The feature dimension is halved after the max pooling layer, and is repeated after each ResBs, reducing spatial dimensions while increasing the feature depth. Each ResB also utilizes skip connections to mitigate the vanishing gradient problem.

we modified the final fully connected (FC) layer to incorporate patient age and  gender. These two features were passed through another FC layer with 10 neurons prior to inclusion in the final layer. We also used a relatively large kernel size of 15 in the first convolutional kernel, and a large kernel size equal to 7 in the latter convolutional kernels. Previous work has shown that large kernel sizes are more helpful for networks to learn meaningful features.\cite{hannun2019cardiologist} \cite{zhao2020adaptive}

\subsubsection{Training}
In our model training, we employed a \textbf{weighted cross-entropy loss function} to address class imbalances. This approach assigns higher weights to underrepresented classes, ensuring the model pays more attention to these samples, thereby improving performance on imbalanced datasets \cite{phan2020resolving}.

To ensure robustness, we conducted \textbf{stratified five-fold cross-validation}\cite{tiittanen2021novel}, maintaining consistent class distribution across folds. Each fold was evaluated using a weighted accuracy metric that penalizes misclassifications based on clinical significance \cite{medtric_metric}.

In our approach, we implemented a \textbf{checkpoint-driven staged training strategy}. While inspired by cyclical approaches such as \textbf{cyclical learning rates (CLR)} \cite{smith2015cyclical}, our method differs significantly. Instead of varying the learning rate continuously within each cycle, we segmented training into stages, each comprising 50 epochs.

At the end of each stage, we saved the best model's state across the 50 epochs as a checkpoint, preserving all learned parameters. For the subsequent stage, training resumed from the best-performing checkpoint of the previous stage. This staged training approach aligns conceptually with methods like early stopping \cite{prechelt1998automatic} and fine-tuning \cite{howard2018universal}, which aim to enhance model generalization and convergence efficiency.

This staged training process was repeated for a total of 5 cycles, effectively allowing the model to fine-tune its parameters iteratively. By restarting from the best checkpoint of the previous cycle, we sought to balance convergence and over-fitting, ensuring optimal utilization of computational resources while improving robustness.

\subsection{Fuzzification}
Fuzzification is a technique we used to refine the predictions of the model and convert numerical confidence scores into more interpretable linguistic categories. This categorization improves interpretability, allowing clinicians to intuitively grasp the model’s predictions. We calculated strength from the generated labels and scores by the predictive model as:

\[
\resizebox{\columnwidth}{!}{$
\text{strength} = \tanh\left((-1)^{\text{label}} \cdot 
\log\left(
\frac{
\left[\max\left(\min(\text{score}, 1-10^{-15}), 10^{-15}\right)\right]^{0.4}}
{1 - \left[\max\left(\min(\text{score}, 1-10^{-15}), 10^{-15}\right)\right]^{0.4}}
\right)\right)
$}
\]

The term ranges are further categorized as:
\vspace{-0.2em}
\[
\text{term\_ranges} =
\begin{cases}
\text{“severe”} & : \; \text{strength} \in [0.9, 1.0], \; \text{label} = 1 \\
\text{“high”} & : \; \text{strength} \in [0.79, 0.9], \; \text{label} = 1 \\
\text{“medium”} & : \; \text{strength} \in [0.0, 0.79], \; \text{label} = \text{any} \\
\text{“low”} & : \; \text{strength} \in [0.79, 0.9], \; \text{label} = 0 \\
\text{“negligible”} & : \; \text{strength} \in [0.9, 1.0], \; \text{label} = 0
\end{cases}
\]
\vspace{-0.4em}

This process provides a human-readable output \(\hat{y}_i\), which is used as the input for the chatbot system, enabling more intuitive and understandable interactions, where \(\hat{y}_i = f(x_i, \theta)\) and \(\hat{y}_i \in \{\text{"negligible"}, \text{"low"}, \text{"medium"}, \text{"high"}, \text{"severe"}\}\)
\subsection{Chatbot Integration}
The chatbot is designed to deliver the model’s predictions in an interactive and explainable format, facilitating clinician understanding and feedback.

\subsubsection{Objectives}
\begin{itemize}
	\item \textbf{Interaction} Provide an intuitive platform for clinicians to query predictions.
	\item \textbf{Explainability} Enhance transparency by explaining model decisions.
	\item \textbf{Feedback} Allow clinicians to provide corrective feedback or annotations, facilitating iterative system enhancement.
\end{itemize}
\subsubsection{Workflow}
\begin{itemize}
    \item \textbf{Input:} The chatbot receives inputs from the fuzzification layer.
    \item \textbf{Response Generation:} It generates responses using a fine-tuned Large Language Model (LLM) based on medical and diagnostic knowledge. This fine-tuning was done with the help of Retrieval Augmented Generation(RAG) on top of pre-trained model Google 1.5 Flash.
    \item \textbf{Evaluation:} Responses are scored on coverage, grounding, and coherence to ensure quality
\end{itemize}
\begin{figure}[H]
\centering
\includegraphics[width=0.9\linewidth, height=6cm]{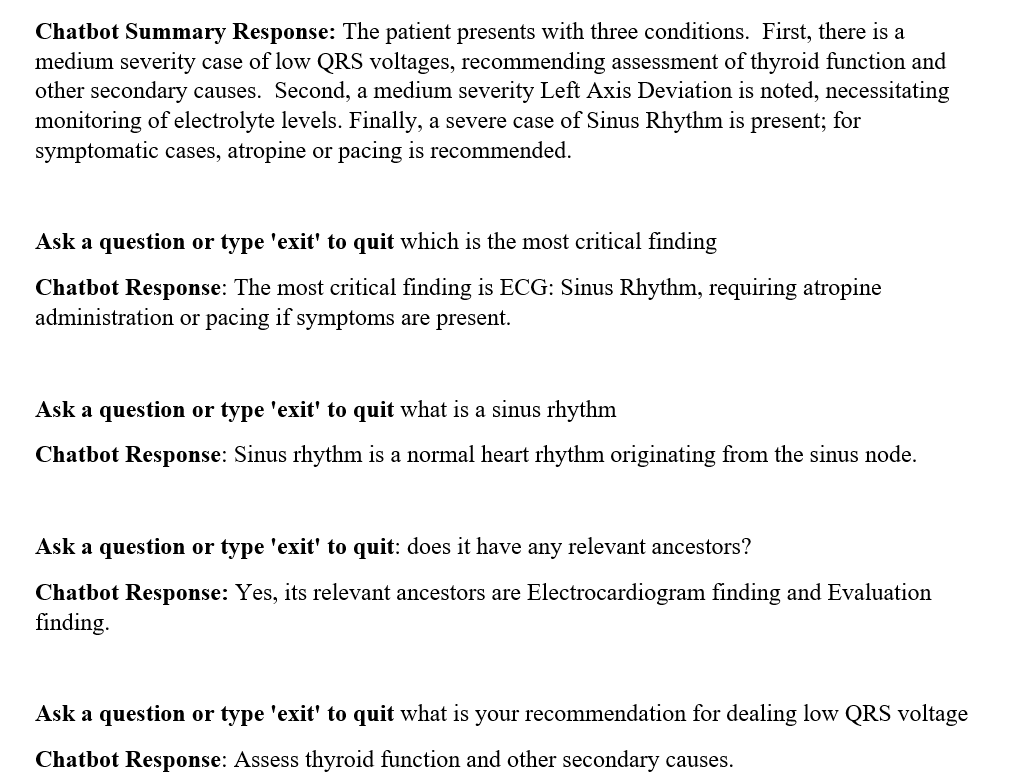}
\caption{Conversation with CardiGPT.}
\label{Fig3:Conversation}
\end{figure}

\section{Evaluation}
The evaluation of Cardi-GPT focuses on two main aspects: the performance of the predictive model and the quality of chatbot responses.

\subsection{Cross-Validation Strategy} 
To validate the performance and generalizability of the model, a stratified five-fold cross-validation strategy was employed. This ensures consistent class distribution across folds, mitigating data imbalance.

During training and validation, all input signals were pre-processed to match the required input size of the convolutional neural network (CNN). Specifically, signals with durations shorter than 4096 samples were zero-padded to maintain uniformity across the dataset. This approach mitigated the effects of variable signal lengths while preserving critical information. More detailed description about this method can be found in \cite{zhao2020adaptive}.

The evaluation dataset consisted of unseen samples, ensuring that the reported performance accurately reflects the model's ability to generalize to new data.

\subsection{Predictive Model Evaluation}

\subsubsection{Evaluation Metric}
We report model performance using a weighted accuracy metric tailored to the clinical context as described in \cite{alday2020classification}. This metric accounts for the varying levels of risk associated with misclassifications, rewarding predictions that misclassify into classes with similar clinical implications. The metric, denoted as \(C\), is defined as:  
\[
C = \frac{S_{\text{observed}} - S_{\text{inactive}}}{S_{\text{correct}} - S_{\text{inactive}}}
\]
where \(S_{\text{observed}}\), \(S_{\text{correct}}\), and \(S_{\text{inactive}}\) represent the observed score, correct score, and inactive score, respectively. The weighted score \(S_P\) is computed as:
\[
S_P = \sum_{j=1}^n \sum_{k=1}^n W_{jk} A_{jk}^P
\]
with \(A_{jk}^P\) defined as:
\[
A_{jk}^P = \frac{1}{N} \sum_{i=1}^N \mathbb{I}(y_{\text{true}}^{(i)} = j \land y_{\text{pred}}^{(i,P)} = k)
\]
Here, \(N\) is the maximum number of class counts, and \(\mathbb{I}\) is the indicator function. Predictions are determined as:
\[
y_{\text{pred}}^{(i)} = \mathbb{I}(\hat{y}_{\text{pred}}^{(i)} \geq \text{threshold})
\]

\subsection{Chatbot Evaluation}
The efficacy of Chatbot depends on the response's relevance to the question asked. A dataset of 1000 question-response pairs based on Chatbot's conversation was created and the relevance of each question-response pair was calculated separately and results were averaged over the entire dataset. Fig. 5. shows 11 such samples and breakdown of their individual relevance scores. The Average- Final Score is calculated over the entire dataset.

\subsubsection{Coverage}
We used NLP tokenization and Large Language Models (LLMs) to identify sub-parts of questions. We also used cosine similarity between sub-part embeddings and the response embedding to determine coverage.

\subsubsection{Grounding}
Specific documents used by the chatbot for response generation were extracted and vectorized. We measured alignment between these documents and the chatbot's response using cosine similarity.
\subsubsection{Coherence}
LLMs pre-trained on grammar evaluation tasks were used to score coherence.

\subsubsection{Final Scoring}
The three metrics were weighted as follows: Coverage (60\%), Grounding (30\%), and Coherence (10\%). The final evaluation score for the chatbot responses was 73\%, reflecting strong overall performance with room for improvement in grounding.

\begin{figure*}[t] 
    \centering
    \begin{subfigure}{0.32\textwidth}
        \centering
        \includegraphics[width=\textwidth]{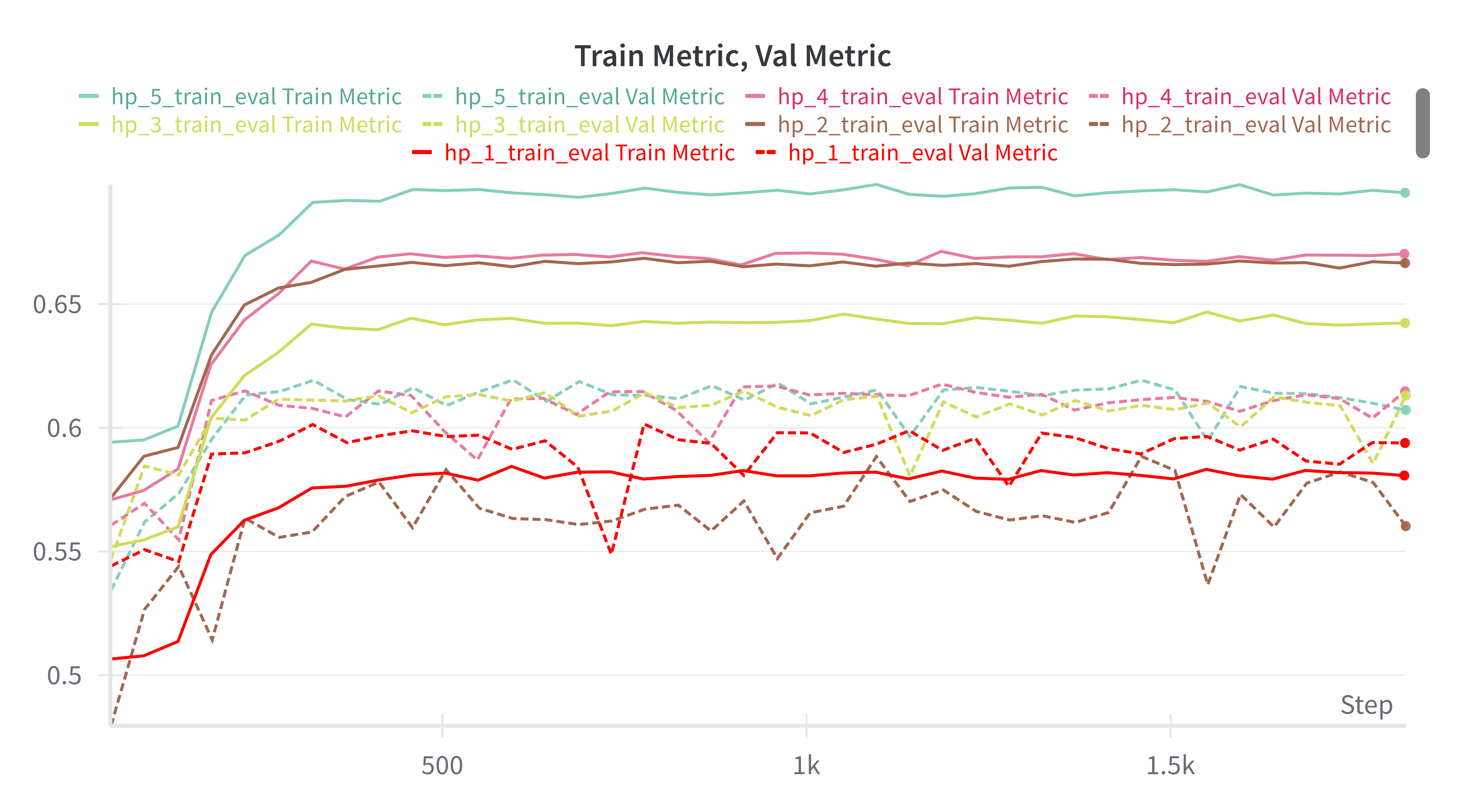}
        \caption{Timestamp-wise Train-Validation scores.}
        \label{Fig3:Timestampwise Train-Validation scores}
    \end{subfigure}
    \hfill 
    \begin{subfigure}{0.32\textwidth}
        \centering
        \includegraphics[width=\textwidth]{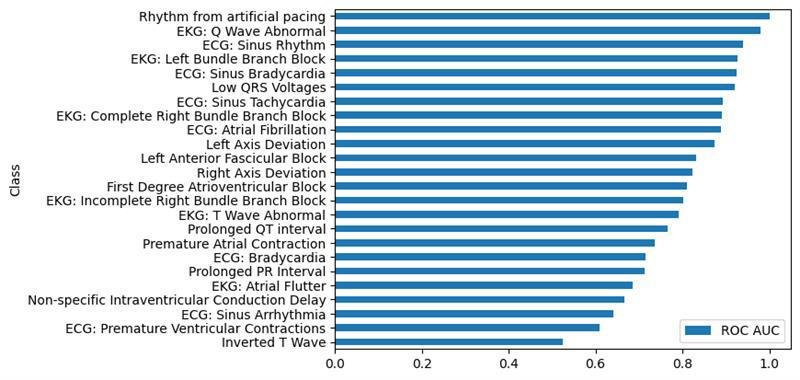}
        \caption{Class-wise prediction scores.}
        \label{Fig3:ROC AUC scores}
    \end{subfigure}
    \hfill 
    \begin{subfigure}{0.32\textwidth}
        \centering
        \includegraphics[width=\textwidth]{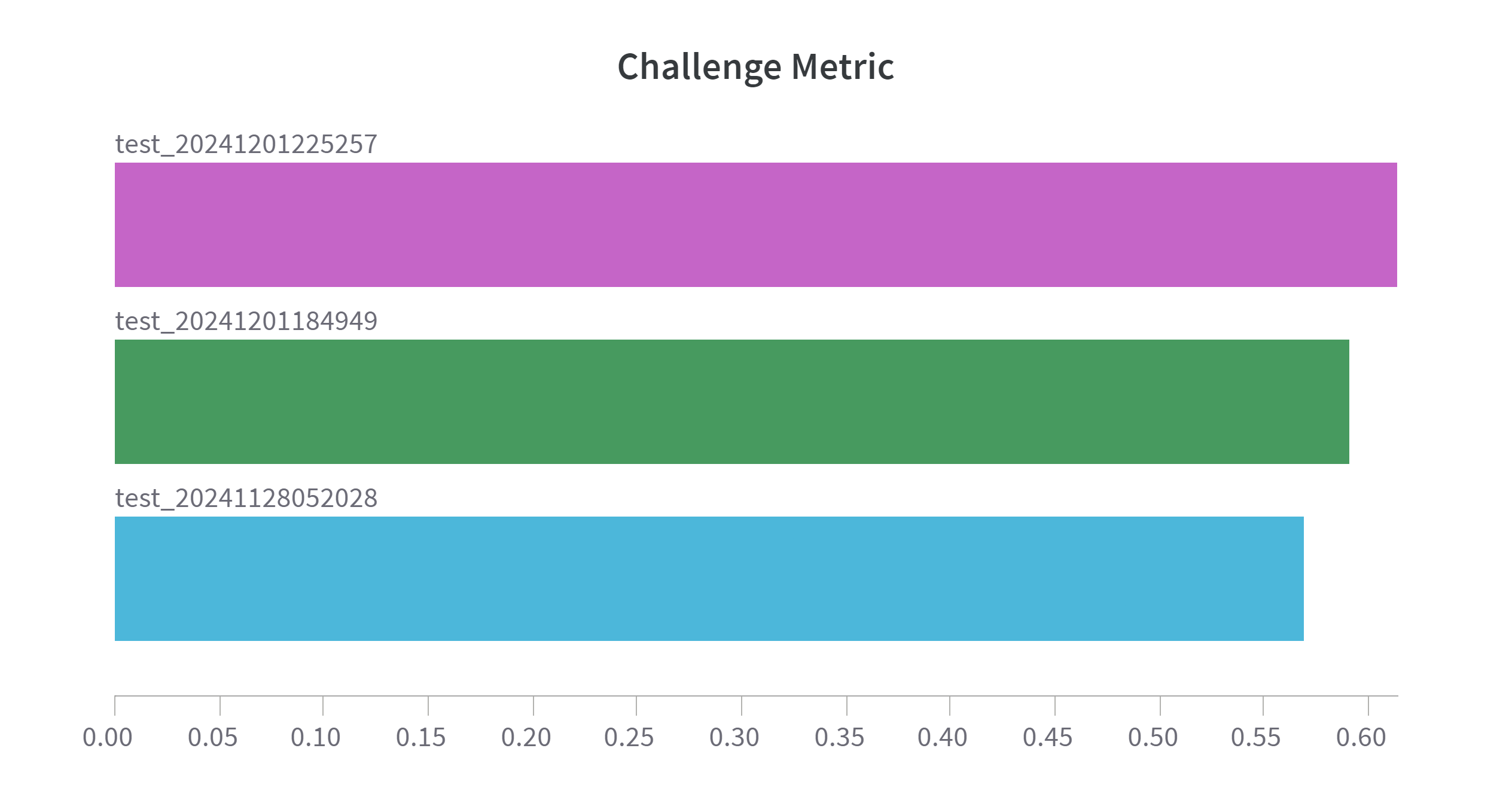}
        \caption{Final Challenge Metric result on Test data.}
        \label{Fig3:Challenge Metric scores}
    \end{subfigure}
    \caption{Visualization of performance metrics}
    \label{fig:combined_figures}
\end{figure*}
\section{Results}

This section presents the outcomes of our evaluation, highlighting the performance of the predictive model and the chatbot system in the Cardi-GPT framework.

\subsection{Predictive Model Performance}
Our model achieved a weighted accuracy score of \(C = 0.6194\) on the evaluation dataset, outperforming some baseline models reported in other studies. For comparison, we used the same ratio of the data into these models that we used for our task. Table~\ref{tab:comparison} provides a comparison of our model's performance against those baseline models:

\begin{table}[h!]
\centering
\caption{Performance Comparison with Models from Other Studies}
\label{tab:comparison}
\begin{tabular}{|l|l|l|}
\hline
\textbf{Model}                 & \textbf{Source Paper} & \textbf{Weighted Accuracy (\(C\))} \\ \hline
Zhao            & \cite{zhao2020adaptive}               & 0.5587                             \\ \hline
Bjorn   & \cite{singstad2020convolutional}               & 0.3468                             \\ \hline
Zhang                  & \cite{zhang2021interpretable}               & 0.5920                             \\ \hline
\textbf{Proposed Model}  & This Study            & \textbf{0.6194}                    \\ \hline
\end{tabular}
\end{table}

The superior performance of the proposed model highlights its ability to learn complex patterns in the data and adapt effectively to the clinical context, making it highly suitable for multi-label classification tasks across 24 condition classes.

\subsection{Fuzzification Results}
The fuzzification process successfully converted numerical confidence scores into linguistic categories, such as "low risk," "moderate risk," and "high risk."
Refinements based on signal strength and inter-lead correlations improved the interpretability of results without sacrificing accuracy.
\subsection{Chatbot System Performance}
\subsubsection{Coverage}
The chatbot scored highly in coverage, addressing over 90\% of the subparts of clinician queries. This demonstrates the system's ability to provide comprehensive answers.

\subsubsection{Grounding}
A grounding score of 70\% was achieved, reflecting a reasonable alignment between chatbot responses and the provided knowledge base. Improvements in training data could further enhance this metric.

\subsubsection{Coherence}
The chatbot exhibited excellent coherence, with responses being grammatically accurate and logically structured. The coherence score was 90\%, indicating strong language generation capabilities.

\subsubsection{Final Evaluation Score}
The combined score for the chatbot system, weighted across coverage (60\%), grounding (30\%), and coherence (10\%), was 73\%.

\subsection{Visualization}

\subsubsection{Predictive Model Performance}
Fig. 5 shows ROC-AUC prediction scores across condition classes.
Range of scores from about 0.5 to nearly 1 indicate strength of classifier in predicting various class categories.
\subsubsection{Chatbot Metrics}

Coverage and grounding metrics were visualized using bar charts, emphasizing strengths and areas needing improvement.
Example chatbot interactions demonstrate the system's ability to handle complex queries with clear, interpretable responses.

\begin{figure}[h]
\centering
\includegraphics[width=0.9\linewidth]{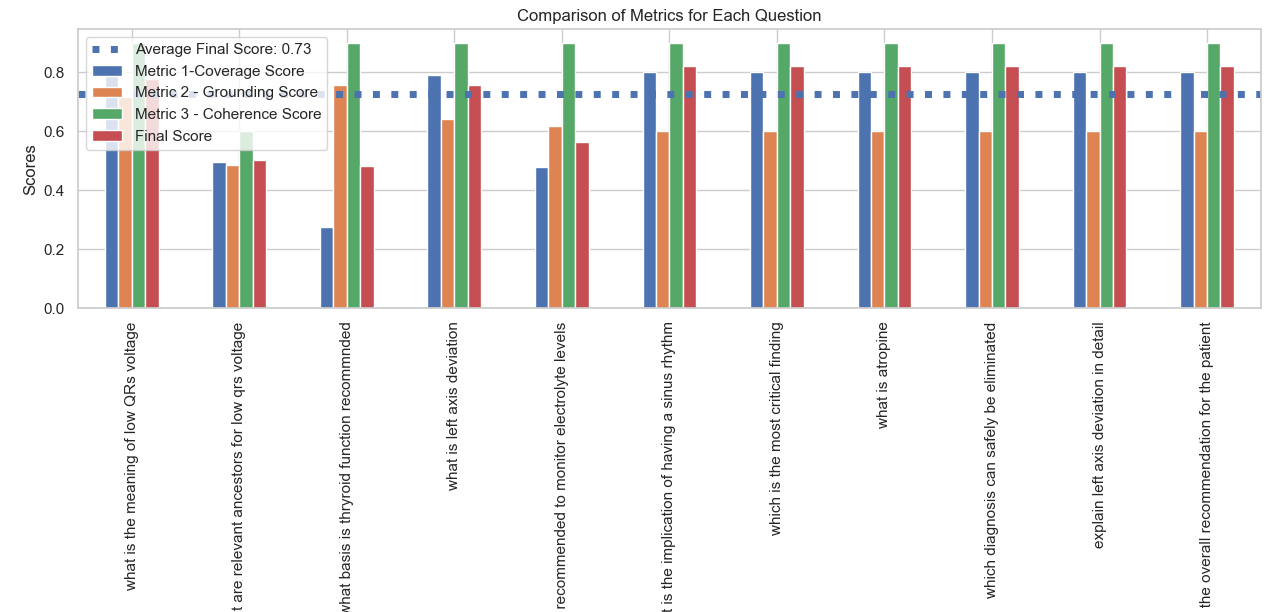}
\caption{Score distribution for CardiGPT conversation.}
\label{Fig3:bargraph scores}
\end{figure}

\begin{figure}[h]
\centering
\includegraphics[width=0.9\linewidth]{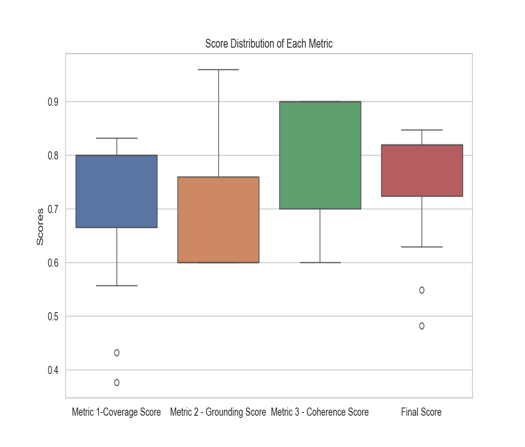}
\caption{Coverage-Grounding-Coherence values feed into Final Score.}
\label{Fig3:boxplot scores}
\end{figure}

\section{Discussion}

The results highlight the potential of Cardi-GPT in addressing ECG interpretation challenges by combining predictive modeling, fuzzification, and interactive explainability.

\subsection{Implications}
\subsubsection{Enhanced Diagnostic Accuracy}
The proposed deep learning model accurately classifies 24 cardiac abnormalities from 12-lead ECGs using a ResNet-SE architecture. The SE layer models inter-channel spatial relationships, achieving a weighted accuracy of 0.6194. The system’s generalization is supported by diverse and heterogeneous training data.

\subsubsection{Improved Interpretability}
Fuzzification translates complex neural network outputs into intuitive insights, enhancing clinician understanding and usability.

\subsubsection{Interactive Decision Support}
The chatbot fosters a collaborative diagnostic process, enabling clinicians to query results, gain explanations, and provide feedback. This improves trust in AI systems and facilitates iterative refinement.

\subsection{Comparison with Existing Systems}
\subsubsection{Traditional Rule-Based Systems}
Static rule-based engines lack adaptability, whereas Cardi-GPT leverages data-driven modeling and feedback integration for dynamic, robust performance.

\subsubsection{Single-Lead Neural Networks}
While single-lead networks focus on simpler data, Cardi-GPT handles the complexity of multi-lead ECGs, offering a more comprehensive diagnostic solution.

\subsubsection{Explainable AI}
Unlike black-box models, Cardi-GPT incorporates explainability through fuzzification and chatbot interaction, making it a clinically practical tool.

\subsection{Limitations}
\subsubsection{Grounding in Chatbot Responses}
The chatbot’s grounding score of 70\% indicates reliance on language generation beyond its knowledge base, posing risks in critical diagnostic contexts.

\subsubsection{Model Generalization}
Although cross-validation shows robustness, performance on imbalanced or noisy datasets warrants further evaluation.

\subsubsection{Fuzzification Thresholds}
Predefined linguistic thresholds in fuzzification may not generalize across diverse populations and require refinement.

\section{Future Work}
\subsubsection{Grounding Improvements}
Enhance chatbot grounding by ensuring responses strictly derive from the provided knowledge base, minimizing potential inaccuracies.

\subsubsection{Adaptive Personalization}
Incorporate patient-specific factors such as comorbidities and medication history to deliver tailored diagnostic insights.

\subsubsection{Integration with Clinical Tools}
Embed Cardi-GPT into electronic health records (EHR) and clinical platforms to streamline diagnostic workflows and improve accessibility.

\section{Conclusion}
This work presents Cardi-GPT, a novel learning-based expert system designed to address the challenges of interpreting 12-lead ECG data. By integrating predictive modeling, fuzzification, and an interactive chatbot system, Cardi-GPT enhances diagnostic accuracy, interpretability, and clinician engagement.

The predictive model achieved a weighted accuracy score of 0.6194, effectively classifying 24 cardiac conditions using expert-verified data. The fuzzification process transformed complex predictions into linguistically interpretable categories, while the chatbot facilitated interactive and explainable decision-making. Together, these components demonstrate the potential of combining neural networks with explainable AI and interactive systems to modernize ECG interpretation.

\bibliography{anthology,custom}
\bibliographystyle{ieeetr}

\end{document}